\documentstyle[12pt]{article}
\title{How to define a unique vacuum in cosmology}
\author{Llu\'{\i}s Bel 
\\
\it Laboratoire de Gravitation et Cosmologie Relativistes\\
\it CNRS/URA 769, Universit\'e Pierre et Marie Curie\\
\it 4, place Jussieu. Tour 22-12. Bo\^{\i}te courrier 142\\
\it 75252 PARIS Cedex 05, France}
\date{}
\begin{document}

\maketitle

\begin{abstract}
We propose a distinguished set of positive and negative energy modes
of the Klein-Gordon equation as a time independent definition of
the vacuum state of a quantized scalar field.
\end{abstract}

a.- {\it Klein-Gordon equation}. Given any space-time with line element:

\begin{equation}
\label {1.1}
ds^2=g_{\alpha\beta}(x^\rho)dx^\alpha dx^\beta, \quad 
\alpha,\beta,\cdots=0,1,2,3
\end{equation}
the Klein-Gordon equation for a classical field $\psi(x^\rho)$ reads,
using a system of units such that $c=1$:

\begin{equation}
\label {1.2}
(\Box-\frac{m^2}{\hbar^2})\psi=(g^{\alpha\beta}\partial_{\alpha\beta}
-\Gamma^\alpha\partial_\alpha-\frac{m^2}{\hbar^2})\psi=0
\end{equation}
where:

\begin{equation}
\label {1.3}
\Gamma^\alpha=g^{\lambda\mu}\Gamma^\alpha_{\lambda\mu}
\end{equation}	
If $\psi_1$ and $\psi_2$ are two, in general complex, solutions then
the current:

\begin{equation}
\label {1.4}
J^\alpha(\psi_1, \psi_2)=
-i\hbar g^{\alpha\beta}(\psi^*_1\partial_\beta\psi_2
-\psi_2\partial_\beta\psi^*_1)
\end{equation}
is conserved:

\begin{equation}
\label {1.5}
\nabla_\alpha J^\alpha
=\frac{1}{\sqrt{-g}}\partial_\alpha(\sqrt{-g}J^\alpha)=0, \quad 
g=det(g_{\alpha\beta})
\end{equation}
and this allows to define the invariant scalar
product $(\psi_1,\psi_2)$ of two well behaved solutions as the flux
of the preceding current across any space-like hypersurface $\Sigma$:

\begin{equation}
\label {1.6}
(\psi_1,\psi_2)=\int_\Sigma{J^\alpha d\Sigma_\alpha}
\end{equation}

b.- {\it Quantization of a scalar field}. The canonical quantization
of a scalar field is a two step process. The first step consists in
selecting a distinguished set of modes of the Klein-Gordon equation
to define what in the jargon of Quantum field theory is called the
{\it Vacuum state}. The second step consists in implementing the
so-called canonical commutation relations to be satisfied by the
field oprator and its conjugate momentum. Only the first step raises
new problems in general relativity and this paper is entirely
dedicated to it. A common belief\,\footnote{See for example
\cite{BD}} is that in general and in particular
in cosmology there is no unique way of choosing a unique vacuum state
and therefore that there is an inevitable spontaneous creation of
particles. We claim in this paper that for Robertson-Walker models
with flat space-sections it is possible to distinguish a preferred
set of modes, with no time mixing of positive and negative modes,
thus defining a unique vacuum state and suppressing the spontaneous
particle creation out from the vacuum state.

In Minkowski space-time and in a galilean frame of reference the
vacuum state is defined by the set of modes $(\epsilon=\pm)$:

\begin{equation}
\label {1.31}
\varphi_\epsilon(x^\alpha, \vec k)=[2\omega(\vec k)(2\pi\hbar)^3]^{-1/2}
u_\epsilon(t,\vec k)e^{\frac{i}{\hbar}\vec k \vec x}, \quad
\varphi_{-\epsilon}(x^\alpha, \vec k)=\varphi_\epsilon^*(x^\alpha, -\vec k)
\end{equation}	
with:

\begin{equation}
\label {1.32}
u_+(t, \vec k)= (2\omega)^{-1/2}e^{-\frac{i}{\hbar}\omega t} \quad 
\omega(\vec k)=+({\vec k}^2+m^2)^{1/2}
\end{equation} 		  
where $\vec k$ is a constant index vector. The modes $\varphi_+$ are
by definition the positive energy modes and $\varphi_-$ the negative
energy modes. This set of modes can be characterized by the following 
conditions:

i) The set of modes, having the general form \ref{1.31}, must be a
complete orthonormal set of particular solutions of the Klein-Gordon
equation. Orthonormality here means that the scalar product of two modes
is:

\begin{equation}
\label {1.33}
(\varphi_{\epsilon_1}(x^\alpha, {\vec k}_1),
\varphi_{\epsilon_2}(x^\alpha, {\vec k}_2))=
\frac{1}{2}(\epsilon_1+\epsilon_2)\delta_{\epsilon_1\epsilon_2}
\delta({\vec k}_1-{\vec k}_2)
\end{equation}
and completeness means that any solution of the Klein-Gordon equation
that can be written as a Fourier transform on the flat space-sections
$t=const$:

\begin{equation}
\label {1.34}
\psi(t, \vec x)=\frac{1}{(2\pi\hbar)^{3/2}}\int	
c(t, \vec k)e^{\frac{i}{\hbar}\vec k \vec x} \, d^3\vec k
\end{equation} 
can also be written as:

\begin{equation}
\label {1.35}
\psi(t, \vec x)=\int	
(a_+(\vec k)\varphi_+(x^\alpha, \vec k)
+a_-(\vec k)\varphi_-(x^\alpha,\vec k)) \, d^3\vec k
\end{equation}

ii) The functions $u_\epsilon$ are solutions of the first order
differential equation:

\begin{equation}
\label {1.36}
i\hbar\dot u_\pm=\pm\omega u_\pm, \quad \dot u=\frac{du}{dt} 
\end{equation}
It is this condition that guarantees that there will not be time mixing of
positive and negative modes. Generalizing to
Roberson-Walker space-times with flat space-sections this second
condition is the main contribution of this paper. 

c.-{\it Robertson-Walker models}. The line-element of a
Robertson-Walker cosmological model with flat space-sections is:

\begin{equation}
\label {1.7}
ds^2=-dt^2+e^{2\sigma(t)}\delta_{ij}dx^i dx^j, \quad
i,j,\cdots=1,2,3
\end{equation}
and the Klein-Gordon equation reads:

\begin{equation}
\label {1.8}
(-\partial_t^2-3\dot{\sigma}\partial_t+e^{-2\sigma}\triangle
-\frac{m^2}{\hbar^2})\psi=0
\end{equation}
where:

\begin{equation}
\label {1.9}
\dot{\sigma}=\frac{d\sigma}{dt}, \quad
\triangle=\delta^{ij}\partial_{ij}
\end{equation}

The scalar product of two solutions can then be written using as
hypersurface $\Sigma$ any space-section $t=const.$:

\begin{equation}
\label {1.10}
(\psi_1,\psi_2)=
i\hbar e^{3\sigma}
\int_t(\psi^*_1\partial_t\psi_2-\psi_2\partial_t\psi^*_1)\, d^3\vec{x}
\end{equation}

d.- {\it Modes}. We shall define a mode, as it is
usual in this case, as a solution of the following form:

\begin{equation}
\label {1.11}
\varphi(x^\alpha,\vec{k})=\frac{1}{(2\pi\hbar)^{3/2}}u(t,\vec{k})e^{\frac{i}{\hbar}\vec{k}\vec{x}}
\end{equation}
where $\vec k$, the index of the mode, is a constant vector and where $u$ must 
therefore be a solution of the following evolution second order
differential equation:

\begin{equation}
\label {1.12}
\hbar^2\ddot u+3\hbar^2\dot\sigma\dot u + \omega^2 u=0, \quad 
\omega^2=e^{-2\sigma}{\vec k}^2+m^2
\end{equation}
The scalar product of two modes corresponding to any two vector
indices is:

\begin{equation}
\label {1.13}
(\varphi(x^\alpha,{\vec k}_1),\varphi(x^\alpha,{\vec k}_2))
=i\hbar
e^{3\sigma}(u^{*}(t,{\vec k}_1)\dot{u}(t,{\vec k}_2)
-u(t,{\vec k}_2){\dot{u}}^*(t,{\vec k}_1)
\delta({\vec k}_1-{\vec k}_2)
\end{equation} 

e.- {\it Reduction of the evolution equation}\,\footnote{The concept
of order reduction has been used widely in many contexts.The use of
this concept here is elementary. Another, non
elementary, application to cosmology can be seen in \cite{BS}}. Let us consider for
each index $\vec k$ the following first order differential equation:

\begin{equation}
\label {1.14}
i\hbar \dot u(t,\vec k)=f(t,\vec k)u(t,\vec k)
\end{equation}
We say that this equation is a reduction of the corresponding
evolution equation \ref{1.12} if the function $f$ is such that every
solution of \ref{1.14} is also a solution of \ref{1.12}. Or
equivalently, if $f$ is such that

\begin{equation}
\label {1.15}
u(t,\vec k)=A(t_0,\vec k)e^{-\frac{i}{\hbar}\int^t_{t_0}f(s,\vec k)\, ds}
\end{equation}  
where $A$ is a constant which will depend on a normalization
condition and on the lower limit of integration that has been chosen,
is a solution of \ref{1.12}. 

Deriving both members of eq. \ref{1.14}, multiplying by $i\hbar$ and
taking into account eq. \ref{1.14} itself we get:

\begin{equation}
\label {1.16}
-\hbar^2\ddot u=i\hbar \dot f u+f^2 u,
\end{equation}
and using eq. \ref{1.12} and dividing by $u$: 

\begin{equation}
\label {1.17}
i\hbar \dot f+f^2+3 i \hbar\dot\sigma f-\omega^2=0
\end{equation}
This is a Riccati equation that $f$ has to satisfy if eq. \ref{1.12}
is to be a reduction of eq. \ref{1.14}.

Let us require that $f$ could be expanded as a Laurent series:

\begin{equation}
\label {1.18}
f=\sum_{n=-s}^{\infty}(i\hbar)^n f_n, \quad s<\infty
\end{equation}	
Substituting this expression into \ref{1.17} we obtain to begin with:

\begin{equation}
\label {1.19}
f_n=0 \quad \hbox{for}\quad -s\le n<0
\end{equation}
and:

\begin{equation}
\label {1.20}
f_0^2=\omega^2
\end{equation}
The following terms are all given by an equation of the following
type:

\begin{equation}
\label {1.21}
f_n=\frac{1}{2f_0}A(f_1,f_2,\cdots,f_{n-1},
\dot f_1,\dot f_2,\cdots,\dot f_{n-1}, t)  
\end{equation}
and can be calculated successively starting with either one of the
solutions of eq. \ref{1.20}. In particular we have

\begin{equation}
\label {1.37}
f_1=-\frac{1}{2}(3\dot\sigma+f^{-1}_0{\dot f}_0)
\end{equation}
and therefore the behavior of \ref{1.15} when $\hbar\rightarrow 0$ is:

\begin{equation}
\label {1.38}
u(t,\vec k)\rightarrow
Ae^{3/2(\sigma(t_0)-\sigma(t))}(f_0(t_0)/f_0(t))^{1/2}
e^{-\frac{i}{\hbar}\int^t_{t_0}f_0(s,\vec k)\, ds}
\end{equation} 
We shall write:

\begin{equation}
\label {1.22}
f_0^{+}=+\omega, \quad f_0^{-}=-\omega, \quad \omega>0
\end{equation}
and note $f_{+}$ and $f_{-}$ the two particular solutions of the
Riccati equation \ref{1.17} that they generate.

The complex conjugate of eq. \ref{1.17} can be written as:

\begin{equation}
\label {1.23}
i\hbar\frac{d\ }{dt}(-f^*)+(-f^*)^2+3i\hbar\dot\sigma(-f^*)-\omega^2=0 
\end{equation}
which proves that if $f$ is a solution of \ref{1.17} then $-f^*$ is
also a solution. Since $f^-_0=-(f^+_0)^*$ because $\omega$ is real, and
since each of these initial terms characterizes the corresponding solution 
$f_-$ and $f_+$ it follows that:

\begin{equation}
\label {1.24}
f_-=-f^*_+
\end{equation} 

Notice that since $\omega$ in \ref{1.12} is a function of ${\vec k}^2$
both functions $f_\pm$ are even functions of $\vec k$:

\begin{equation}
\label {1.31x}
f(t,\vec k)=f(t,-\vec k)
\end{equation} 

f.- {\it Positive and negative energy modes}. We shall define the
positive (respectively negative) energy modes $u_+$ (resp. $u_-$) as those 
modes for which $u$
is a solution of \ref{1.14}, $f$ being $f_+$ (respectively $f_-$):

\begin{equation}
\label {1.46}
i\hbar{\dot u}_\pm(t,\vec k)=f_\pm(t,\vec k)u_\pm(t,\vec k)
\end{equation}
or:
\begin{equation}
\label {1.39}
u_\pm(t,\vec k)=A_\pm(t_0,\vec k)
e^{-\frac{i}{\hbar}\int^t_{t_0}f_\pm(s,\vec k)\, ds}
\end{equation}

Let $c(t,\vec k)$ be a solution of the second order equation \ref{1.12}
corresponding to initial conditions $c(t_0,\vec k)$ and 
$\dot c(t_0,\vec k)$ on some space-section $t=t_0$. If $u_+$ and
$u_-$ are two particular energy modes, one positive and the other
negative there will exist two constants $a_\pm(\vec k)$ such that:

\begin{equation}
\label {1.42}
c(t_0,\vec k)=a_+(\vec k)u_+(t_0,\vec k)+a_-(\vec k)u_-(t_0,\vec k)
\end{equation}
and from \ref{1.14} we shall have also:

\begin{equation}
\label {1.43}
i\hbar\dot c(t_0,\vec k)=a_+(\vec k)f_+(t_0,\vec k)u_+(t_0,\vec k)
+a_-(\vec k)f_-(t_0,\vec k)u_-(t_0,\vec k)
\end{equation}
Solving for $a_\pm$ we obtain:

\begin{equation}
\label {1.44}
a_\pm(\vec k)=\frac{i\hbar\dot c(t_0,\vec k)
-f_\mp(t_0,\vec k)c(t_0,\vec k)}{u_\pm(t_0,\vec k)
(f_\pm(t_0,\vec k)-f_\mp(t_0,\vec k))}
\end{equation}
If $c$ is itself a positive (resp. negatif) energy mode  then $a_+=0$
(resp. $a_-=0$) demonstrating explicitly that an energy mode is
positive (resp. negative) independently of the choice of the
space-section and initial condition on it, as far as it satisfies the
appropriate first order equation \ref{1.46}.

From \ref{1.14} and \ref{1.24} we have:

\begin{equation}
\label {1.40}
i\hbar\frac{d }{dt}(u_+-u^*_-)=f_+(u_+-u^*_-)
\end{equation}
and therefore we shall have:

\begin{equation}
\label {1.41}
u_-(t,\vec k)=u^*_+(t,\vec k)
\end{equation}
provided that we choose initial conditions which satisfy this
condition on some arbitrary space-section.
From \ref{1.31x} we shall have:
 
\begin{equation}
\label {1.32}
u_\pm(t,\vec k)=u_\pm(t,-\vec k)
\end{equation} 

Let us consider two modes with energy condition $\epsilon_1=\pm$ and
$\epsilon_2=\pm$. From \ref{1.13}, and \ref{1.14} and its complex
conjugate, we have:

\begin{equation}
\label {1.25}
(\varphi_{\epsilon_1}(x^\alpha,{\vec k}_1),
\varphi_{\epsilon_2}(x^\alpha,{\vec k}_2))=  
e^{3\sigma}(u^{*}_{\epsilon_1}(t,{\vec k}_1)u_{\epsilon_2}(t,{\vec k}_2)
(f_{\epsilon_2}(t,{\vec k}_2)+f^*_{\epsilon_1}(t,{\vec k}_1))
\delta({\vec k}_1-{\vec k}_2)
\end{equation}
From this result we can see that the scalar product of two different
modes is zero. If ${\vec k}_1 \not={\vec k}_2$ then this follows
from the properties of the Dirac $\delta$ function. 
If ${\vec k}_1={\vec k}_2$ and $\epsilon_1=-\epsilon_2$ then this is a
consequence of \ref{1.24}. To normalize the modes we shall require
the generalized modes to satisfy again conditions \ref{1.33}.
This is equivalent to requiring the constant $A_\pm(t_0,\vec k)$ in \ref{1.15} 
to satisfy:

\begin{equation}
\label {1.27}
A_\pm A^*_\pm(t_0\vec k)=e^{-3\sigma}
{\mid f_\pm(t_0,\vec k)+f^*_\pm(t_0,\vec k)\mid}^{-1}
\end{equation}
This fixes the norm of $A_\pm$ but not its phase. Since we want to have
everywhere the relation \ref{1.41} we may require that $A_\pm$ satisfy
the relation:

\begin{equation}
\label {1.45}
A_-(t_0,\vec k)=A^*_+(t_0,\vec k)
\end{equation}

Let $\psi$ be any solution of the Klein-Gordon equation \ref{1.8}
which can be expressed as a Fourier integral on each space-section
$t=constant$: 

\begin{equation}
\label {1.28}
\psi(t, \vec x)=\frac{1}{(2\pi\hbar)^{3/2}}
\int c(t,\vec k)e^{\frac{i}{\hbar}\vec k\vec x}\, d^3\vec k 
\end{equation}
Since the $c$'s must be, for each $\vec k$, a solution of \ref{1.12}
they will be a linear combination of $u_+$ and $u_-$ as in \ref{1.42}.
Substituting in \ref{1.28}, using \ref{1.32}, and after some re-writing we get:

\begin{equation}
\label {1.30}
\psi(t, \vec x)=
\int a_+(\vec k)\varphi_+(t,\vec k)
+a_-(\vec k)\varphi_-(t,\vec k) \, d^3\vec k 
\end{equation}
where:
\begin{equation}
\label{1.47}
\varphi_-(t,\vec k)=\varphi^*_+(t,-\vec k)
\end{equation}

We have thus proved that the set of modes defined above provide a
straightforward generalization of the modes \ref{1.31} associated with
the galilean frames of reference of Minkowski space-time. The vacuum
state that they define does not depend on time.

g.- {\it Concluding remarks}. We have assumed that the space-sections
$t=constant$ were flat. This is not an essential restriction. Models
with non flat space-sections can be dealt with using the
eigen-states of the Laplacian of a constant curvature 3-dimensional
riemannian metric\,\footnote{See for instance \cite{LK}}.

Also, the main idea of this paper which consisted in reducing the
Klein-Gordon equation to two complex conjugate first order equations
with respect to time can be generalized to more
general space-times and frames of reference. This work will be
published elsewhere.

\end{document}